# Detection of electrical spin injection by light-emitting diodes in top- and side-emission configuration

R. Fiederling, P. Grabs, W. Ossau, G. Schmidt, and L.W. Molenkamp

*Physikalisches Institut (EPIII), Würzburg University, Würzburg, Germany*

Detection of the degree of circular polarization of the electroluminescence of a light-emitting diode fitted with a spin injecting contact (a spin-LED) allows for a direct determination of the spin polarization of the injected carriers. Here, we compare the detection efficiency of (Al,Ga)As spin-LEDs fitted with a (Zn,Be,Mn)Se spin injector in top- and side-emission configuration. In contrast with top emission, we cannot detect the electrical spin injection in side emission from analysing the degree of circular polarization of the electroluminescence. To reduce resonant optical pumping of quantum-well excitons in the side emission, we have analysed structures with mesa sizes as small as 1 μm.

The efficient electrical injection and detection of a spin polarized current into semiconductor heterostructures is a key element for the development of semiconductor spintronics [1, 2]. Originally, much effort was devoted to detecting magnetoresistance changes in ferromagnet / semiconductor / ferromagnet interfaces [3, 4], where, because of the very small effects observed, it turned out to be difficult to find evidence for spin injection [5]. An important step forward could only be made when zincblende-based light-emitting diodes (LED) (e.g. in the (Al,Ga)As system) were used as detectors for electrical spin injection. In this scheme, already proposed in the 70's by Aranov and Pikus [6], the spin polarized current is converted into circularly polarized electroluminescence; the degree of circular polarization of the electroluminescence of the spin-LED is directly proportional to the spin polarization of the carriers in the detection quantum well (QW). This approach has yielded evidence for spin injection using semi- [7] or ferro-magnetic [8] semiconductors, as well as ferromagnetic metal [9, 10, 11] contacts on GaAs based LEDs.

Spin-LEDs are typically used in a top-emission measurement-geometry, where the external magnetic field is parallel to the current path, and parallel to the wave vector of the emitted light [7, 9, 10, 12, 13], as depicted in Fig. 1(a). However, for ferromagnetic materials like (Ga,Mn)As where the magnetic moment of the manganese system is typically perpendicular to the growth axis, side-emitted electroluminescence has been used to infer the spin polarization [8]. In this geometry the magnetic field is also parallel to the wave vector of the emitted light, but at the same time perpendicular to the growth axis of the structure and the confinement axis of the QW (see Fig. 1(a)). This quasi-Voigt geometry gives rise to different selection rules for optical transitions in GaAs based QWs [14] compared with those applicable for top emission [15]. If the selection rules are strictly valid, they may prohibit the determination of electron spin polarization by circular polarized light. Recent work on (Ga,Mn)As based spin-LEDs reveals a remarkable influence of the measurement geometry on the observed spacer layer thickness dependence of the circular polarization of the electroluminescence [16].

In view of this, it seems of importance to investigate the efficiency of spin injection in both emission configurations for II-VI-based spin-LEDs, which demonstrated the most efficient spin injection observed so far [7]. Here, we address this issue. We study spin-LEDs grown by MBE, using two separate growth chambers for the III-V LED and the II-VI spin aligner layer, which are connected by an UHV transfer system. The LED is grown on p-type GaAs substrate and GaAs p-type buffer layer followed by a 500 nm $Ga_{0.96}Al_{0.04}As$ p-type ($1\times10^{19}$ cm$^{-3}$) barrier, a 25 nm undoped $Ga_{0.96}Al_{0.04}As$ spacer, a 15 nm GaAs quantum well (QW), again a 5 nm undoped $Ga_{0.96}Al_{0.04}As$ spacer, followed by an 100 nm n-type ($2\times10^{16}$ cm$^{-3}$) barrier layer. The 300 nm n-type ($2\times10^{18}$ cm$^{-3}$) doped spin aligner layer ($Be_{0.08}Zn_{0.87}Mn_{0.05}Se$) is grown on top of the LED followed by a 20 nm heavily n-type doped ($1\times10^{19}$ cm$^{-3}$) ZnSe layer which improves the contacting properties. The structure is covered in situ by an Al contact layer. Using standard optical lithography, three different types of LED structures A, B, and C were fabricated.

Structure A was patterned into 500 µm x 500 µm mesas fitted with 100 µm x 100 µm top contacts, while structure B and C were patterned into stripes with a length of 350 µm and a width of $d_{LED}$=5 µm and $d_{LED}$=1 µm, respectively. Structure B and C were contacted through 160 µm x 400 µm bond pads placed at the ends of the stripes (see Fig. 1(b)).

The samples were cooled down to 1.6 K in a liquid helium bath optical cryostat fitted with a split-coil magnet, so that measurement configuration could be varied from Faraday to quasi-Voigt geometry without changing the optical path of the light. The electroluminescence (EL) was focused on the entrance slit of a 1m monochromator and detected by a cooled CCD. The samples could be additionally illuminated by an Ar-ion laser with a wavelength of 514 nm, so that photoluminescence (PL) of the GaAs QW could be recorded in the same setup geometry as for the EL. The degree of circular polarization of the EL or the PL was determined by a λ/4 plate followed by a linear polariser. Note that the spin polarization of the electrons injected into the GaAs LED is equal in both measurement configurations, since the magnetisation of the spin aligner layer is isotropic. This is confirmed by our recent transport experiments on spin injection [17], where we obtain identical magneto-resistance effects for in- and perpendicular to plane magnetic field configuration.

In a first series of experiments, the electrical spin injection was measured for sample A, driving the spin-LED with a current of 5 mA. EL spectra of the GaAs QW recorded at $B$=0 and $T$=1.6 K are shown in Fig. 2(a). All spectra exhibit two features, identified as exciton (X) and bound exciton (BX) recombination, with an energy separation of 2.9 meV and a linewidth of 2.4 meV. The energy position of the exciton (1.5254 eV) corresponds to a GaAs QW of 15 nm width with 4 % Al content in the barrier layer. Due to the selection rules for optical transitions in GaAs QWs in Faraday configuration, the optical polarization degree is equal to the spin polarization of the electrons in the detection QW [15]. The optical polarization degree is given by $P_C = (I^{\sigma+} - I^{\sigma-})/(I^{\sigma+} + I^{\sigma-})$ where $I^{\sigma+}(I^{\sigma-})$ are the intensities detected with right (left) hand circular polarization, $\sigma^+(\sigma^-)$, respectively. In Fig. 2(b) EL spectra, recorded at $B$=0.5 T in top and side emission for $\sigma^+(\sigma^-)$ detection are shown for sample A. It is clearly observable that while the top emission signal exhibits already a strong circular polarization, the $P_C$ obtained in side emission is zero. The optical polarization degree of sample A in dependence of the external magnetic field is shown in Fig. 3, detected in the top- (solid squares) and side- (half filled squares) emission configuration. The polarization degree and so the electron spin polarization of sample A (top emission) strongly increases with increasing magnetic field, and reaches a maximum of 0.68 at $B$~2 T, and then slightly decreases for higher magnetic fields. However, in the side emission configuration the circular polarization degree is negligibly small over the entire range of the magnetic field. As a reference for the electrical spin injection signal, we analysed the PL of the GaAs QW. The intrinsic polarization degree of the GaAs QW is displayed in Fig. 3 (stars) and shows small positive values for $B$<3 T, turning negative to $P_C$=-0.05 for $B$=5.5 T. This behaviour of the intrinsic circular polarization is typical for an (Al,Ga)As QW with the dimensions used by us. The PL data thus provides evidence, that the (Be,Zn,Mn)Se layer on top of the structure does not influence the carriers in the GaAs p-i-n, e.g. by means of exchange interactions, and that the light emitted by the QW passing through the II-VI compound does not change its polarization due to circular dichroism effects in the (Be,Zn,Mn)Se layer.

A first conclusion that directly can be drawn from our EL data is that despite the high-efficiency of the electrical spin injection into the LED, as evidenced by the $P_C$ in Faraday configuration, almost no circularly polarized light emission is obtained in the quasi-Voigt configuration. This implies that it is questionable to use the quasi-Voigt configuration for any evidence of electrical spin injection.

Several effects can be responsible for the poor spin polarization detection quality in side emission configuration; The optical selection rules for heavy hole transitions in principle prohibit the emission of circular polarized light along the QW planes [14]. Nevertheless, the selection rules under additional applied electric and magnetic fields are expected to relax, so that detection of some circularly polarized EL should be conceivable. Additionally, almost all photons emitted sideways (Sample A) have to travel through at least several hundred microns of III-V material, before leaving the structure. This might lead to multiple reflections at the QW- or at the II-VI-interface planes, destroying the original emitted polarization of the light. In addition, this may also cause resonant pumping of QW excitons along the QW planes, where every absorption and reemission process leads to a reduction of the original emitted circular polarization, due to additional electron spin relaxation.

To address these issues, we have determined the polarization degree for sample B and C (see Fig. 1(b)), along the direction of the short axis of the LED stripes with width of $d_{LED}$=5 µm and $d_{LED}$=1 µm, respectively. Compared to sample A, any kind of resonant pumping, wave guide-, or anisotropic spin relaxation effects in these structures should be strongly reduced. The data for these two samples is presented in Fig. 3. Obviously, the electron spin polarization in top emission for sample B (solid circles) and C (solid diamonds) is smaller than for sample A. This is due to the strong increase of the current density, necessary to drive LED B and C to obtain reasonable intensities for the light emission ($I$=9 mA); the effect will be discussed in detail elsewhere. Even though the electron spin polarization in top emission for sample B and C is still very high, we obtain no evidence for electron spin injection via circularly polarized light emission for sample B and sample C, half filled circles and -diamonds in Fig. 3. Despite of the reduction of the mesa size down to the range of the penetration depth of the emitted light, no evidence of electrical spin injection can be found in side emission configuration.

In conclusion, we have demonstrated that for typical spin-LED structures, the spin detection efficiency for top- and side-emission is drastically different, so that at least in our experiment the side emission cannot be regarded as a reliable indicator for spin-polarized electron injection. We can exclude wave guide effects and resonant pumping in the side emission configuration as a major effect for the reduction of circular polarized light emission, because even for mesas as small as 1 µm, no evidence for electrical spin injection was found. It is more likely that the selection rules for side emission configuration prohibit the measure of the electrical spin injection in our devices.

We acknowledge financial support by the DFG(SFB 410), BMBF, the EC, and the DARPA SPINS programme.

List of figures

Fig. 1.   (a) Sketch of a standard LED structure with a 500 µm mesa. The measurement configurations are either top emission with the light propagating along the growth axis, or side emission, where the light propagating parallel to the QW planes is detected. In both cases the external magnetic field is parallel to the light propagation.
(b) Top view of the LED stripes structures with a mesa width of $d_{LED}$=1 µm or $d_{LED}$=5 µm. Light emission from the bond pad is blocked by black carton, covering the complete bond pad.

Fig. 2.   (a) Electroluminescence signal of the LED in top and side emission configuration at $B$=0 and $T$=1.6 K.
(b) An external magnetic field of $B$=0.5 T results in circular polarized light emission ($\sigma^+/\sigma^-$) in top emission configuration, while the side emission remains unpolarized.

Fig. 3.   The degree of circular polarization ($P_C$) obtained by the electroluminescence of the LED structures with mesa sizes of 500 µm, 5 µm and 1 µm, for top (TE) and side emission (SE) configuration. All top emission EL exhibit a strong $P_C$, while there is no circular polarized light emission in the side emission. The $P_C$ of the photoluminescence from the GaAs QW is the reference for the electrical spin injection.

Figure 1 (Fiederling et al.)

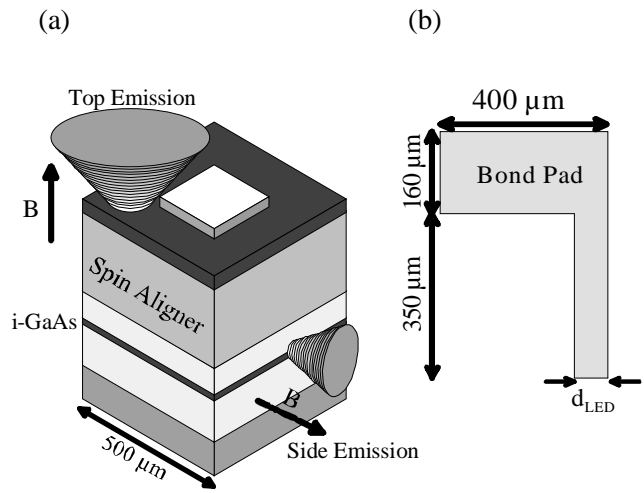

(a)        (b)

Figure 2 (Fiederling et al.)

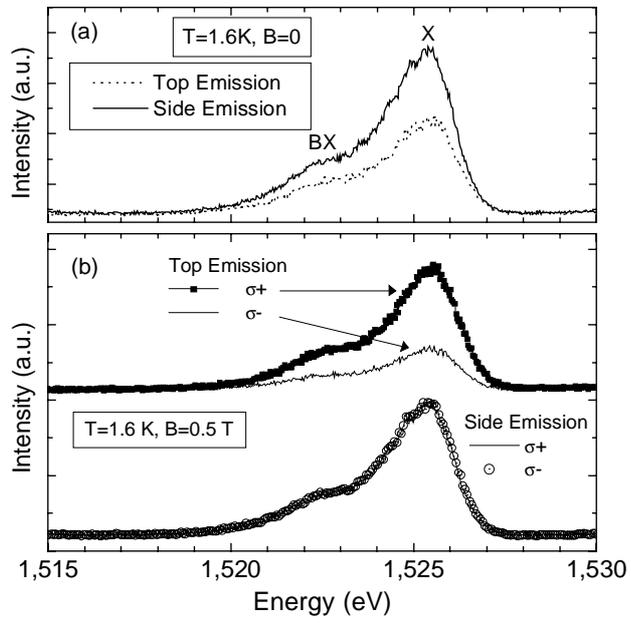

Figure 3 (Fiederling et al.)

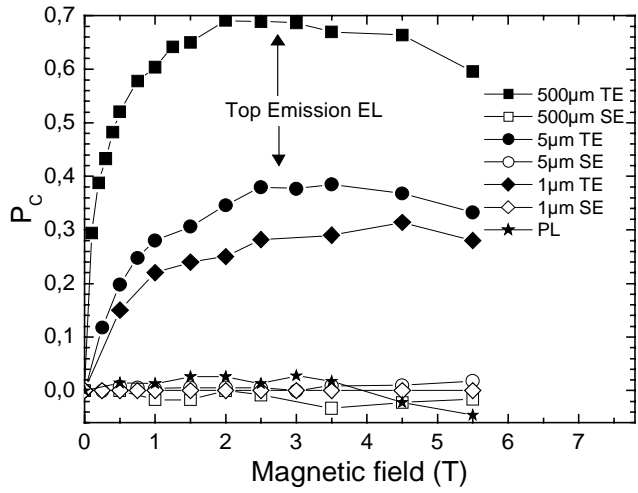

References

1 S. Datta, B. Das, Appl. Phys. Lett. **56** (7), 665 (1990).

2 G. A. Prinz, Physics Today, **48** (4), 58-63 (1995).

3 F. G. Monzon, M. L. Roukes, JMMM 198-199, 632 (1999).

4 P. R. Hammar, B. R. Bennet, M. J. Yang, M. Johnson, Phys. Rev. Lett. **83** (1), 203-206 (1999).

5 G. Schmidt, D. Ferrand, L. W. Molenkamp, A. T. Filip, B. J. van Wees, Phys. Rev. B **62** (8), R4790 (2000).

6 A. G. Aranov, G. E. Pikus, Sov. Phys. Semicond. **10**, 698-700 (1976).

7 R. Fiederling, M. Keim, G. Reuscher, W. Ossau, G. Schmidt, A. Waag, L. W. Molenkamp, Nature (London) **402**, 787 (1999).

8 Y. Ohno, D. K. Young, B. Beschoten, F. Matsukura, H. Ohno, D. D. Awschalom, Nature (London) **402**, 790 (1999).

9 H. J. Zhu, M. Ramsteiner, H. Kostial, M. Wassermeier, H. P. Schönherr, K. H. Ploog, Phys. Rev. Lett., **87** (1), 016601 (2001).

10 A. T. Hanbicki, B. T. Jonker, G. Itskos, G. Kioseoglou, A. Petrou, Appl. Phys. Lett., **80** (7), 1240 (2002).

11 V. F. Motsnyi, J. De Boeck, J. Das, W. Van Roy, G. Borghs, E. Goovaerts, V. I. Safarov, Appl. Phys. Lett., **81** (2), 265 (2002).

12 B. T. Jonker, Y. D. Park, B. R. Bennet, H. D. Cheong, G. Kioseoglou, A. Petrou, Phys. Rev. B, **62** (12), 8180 (2000).

13 T. Gruber, M. Keim, R. Fiederling, G. Reuscher, W. Ossau, G. Schmidt, A. Waag, L. W. Molenkamp, Appl. Phys. Lett. **78** (8), 1101 (2001).

14 Optical Orientation, ed. by F. Meier and B. P. Zakharchenya, Elsevier Science Publishers B. V. (1984). See chap. 2 (5.1). The discussion of strained semiconductors is similar to QW heterostructures.

15 C. Weisbuch & B. Vinter, Quantum Semiconductor Structures – Fundamentals and Applications, Academic, Boston (1991). See chap. 11.

16 D. K. Young, E. Johnston-Halperin, D. D. Awschalom, Y. Ohno, H. Ohno, Appl. Phys. Lett., **80** (9), 1598 (2002).

17 G. Schmidt, G. Richter, P. Grabs, C. Gould, D. Ferrand, L.W. Molenkamp, Phys. Rev. Lett., **87** (22), 227203 (2001).